\documentclass[aps,superscriptaddress,prl,twocolumn,secnumarabic,nobalancelastpage,amsmath,amssymb,nofootinbib]{revtex4}

\usepackage{graphics}      
\usepackage[pdftex]{graphicx,color}
\usepackage{longtable}     
\usepackage{url}           
\usepackage{bm}            

\usepackage{amsmath}
\usepackage{amssymb}
\usepackage{subfigure}

\newcommand\bra[1]{\left<#1\right|}
\newcommand\ket[1]{\left|#1\right>}

\begin{document}

\title{Matter Wave Scattering from Ultracold Atoms in an Optical Lattice}
\author{Scott N. Sanders}
\affiliation{Massachusetts Institute of Technology, Cambridge,
  Massachusetts 02139}
\affiliation{Harvard University, Cambridge,
  Massachusetts 02138}
\email{ssanders@post.harvard.edu}
\author{Florian Mintert}
\affiliation{Albert-Ludwigs-Universit\"{a}t Freiburg,
  Hermann-Herder-Str. 3, 79104 Freiburg}
\author{Eric J. Heller}
\affiliation{Harvard University, Cambridge,
  Massachusetts 02138}
\date{\today}

\begin{abstract}
  We study matter wave scattering from an ultracold, many body atomic
  system trapped in an optical lattice. We determine the angular cross
  section that a matter wave probe sees and show that it is strongly
  affected by the many body phase, superfluid or Mott insulator, of
  the target lattice.  We determine these cross sections analytically
  in the first Born approximation, and we examine the variation at
  intermediate points in the phase transition by numerically
  diagonalizing the Bose Hubbard Hamiltonian for a small lattice.  We
  show that matter wave scattering offers a convenient method for
  non-destructively probing the quantum many body phase transition of
  atoms in an optical lattice.
\end{abstract}

\maketitle

Ultracold atoms in an optical lattice form an optical crystal that
permits the clean implementation of fundamental models of condensed
matter physics to a degree unheard of in other systems. The physical
parameters that characterize this system, namely the depth of the
lattice and the interactions of the atoms in the lattice, are
experimentally tunable so that it is possible to probe the atoms'
behavior controllably from non-interacting single particle mechanics
to strongly interacting, many body
physics~\cite{Zoller1998,Bloch2008,Yukalov2009}. Of particular
interest are the many body theories of quantum phase transitions,
whose predictions can be observed in such a system. 

The superfluid to Mott insulator phase transition has been probed
experimentally by studying the interference of clouds of atoms
expanding freely out of an optical lattice~\cite{Bloch2002}. This
method provides information about the correlation properties of the
ground state of the cold atom system, but does not depend on the
excitation spectrum of the lattice, which is essential to study the
superfluid fraction~\cite{Burnett2003}. Bragg spectroscopy has been
combined with this method in order to examine the excitation spectrum
~\cite{Esslinger2004}; nevertheless, it requires the destruction of
the sample under examination.  More recently, it has been proposed
theoretically that the on-site number statistics can be probed by
observing light scattered into an optical
cavity~\cite{Ritsch2007}. Very recently, light scattering from optical
lattice systems has been analyzed to determine the effect of the many
body phase on the scattering cross section of a photon due to
interactions with a lattice~\cite{Morigi2009}. Very little attention
has been paid to the interaction of matter waves with these optical
crystals. Theoretical work has been done emphasizing the impact of
disordered atoms in a periodic potential on scattering of photons and
particles~\cite{Akulin2001}. We are interested, however, in the impact
of the many body phase on matter wave scattering.

The scattering of matter waves from a lattice system provides a very
useful technique to probe the many body phase transition both because
it does not require the destruction of the lattice under examination
and because it depends strongly on the excitation spectrum of the
target. This is critical to probe superfluidity and the dynamics of
the lattice, beyond analysis of ground state properties. In addition,
the simple form of the interaction between a slow-moving probe atom
and the atoms in the lattice emphasizes structure that depends on the
many body properties of the lattice. As we will show, the differential
cross section is a highly suitable quantity for the distinction
between the Mott insulating and superfluid phases.

Our objective has been to study the scattering patterns of matter
waves due to passage through ultracold atoms trapped in a uniform
optical lattice. The scattering target is well modeled by the Bose
Hubbard Hamiltonian, $H_{\text{BH}}$~\cite{Zoller1998},
\begin{equation}
  \hat{H}_{\text{BH}} = -J \sum_{\left< \bm{R}, \bm{R}^{\prime} \right>}
  \hat{a}_{\bm{R}}^{\dagger} \hat{a}_{\bm{R^{\prime}}} + \frac{1}{2} U
  \sum_{\bm{R}} \hat{n}_{\bm{R}}(\hat{n}_{\bm{R}}-1) 
\end{equation}
The probe is a free particle of mass, m, with Hamiltonian, $H_P =
\hat{p}^2/2m$, which does not interact with the confining light of the
lattice. We require low-energy probes that avoid interband excitations
of atoms in the lattice. In this case, s-wave scattering is dominant,
and we may treat the interaction between the probe and each lattice
atom as a pseudopotential with scattering length, $a_s$. The total
interaction potential is $\hat{V} = \sum_j \frac{2 \pi \hbar^2}{m} a_s
\delta(\hat{\bm{r}}-\hat{\bm{r}}_j)$ \cite{Wodkiewicz1991}. The
operators $\hat{\bm{r}}$ and $\hat{\bm{r}}_j$ give the positions of
the probe and the $j^{\text{th}}$ lattice atom, respectively. The full
Hamiltonian for the scattering interaction is $\hat{H} =
\hat{H}_{\text{P}} + \hat{H}_{\text{BH}} + \hat{V}$.

The states of the lattice before and after scattering of a probe atom
are many body states of the $N$ atoms in the lattice. The particular
ground state in the lattice depends on the relative sizes of the the
interaction strength, $U$, and the tunneling matrix element, $J$,
appearing in the Bose Hubbard model. For weak repulsion between the
atoms in the lattice, the atoms will delocalize and the superfluid
fraction will increase to one as the interaction strength goes to
zero~\cite{Burnett2003}. As the repulsion between the atoms in the
lattice becomes large compared to the tunneling matrix element, the
atoms will localize, the superfluid fraction will go to zero, and a
gap will open in the excitation spectrum, giving rise to the Mott
insulator state. It is possible to alter the interaction strength
between the atoms in the lattice by adjusting the depth of the
lattice, or by manipulating the scattering length of lattice atom
collisions through a Feshbach resonance. It is best for the purpose of
probing the many body phase of the lattice to retain a constant
lattice depth so that the scattering patterns are not trivially
affected by the changing density profile associated with a changing
lattice potential.


The incident probe wavefunction is in the plane wave state,
$\left|\bm{k}_0\right>$, with energy, $\hbar^2 k_0^2/(2m)$. The target
is initially in the ground state, $\ket{n_0}$, of the Bose Hubbard
Hamiltonian, with energy, $E_{n_0}$. We consider the cross section for
a transfer of momentum $\bm{\kappa} = \bm{k_0}-\bm{k}$ from the probe
to the lattice. This is associated with a change of the state of the
atoms in the lattice to a potentially excited state, $\ket{n}$. The
energy gained by the lattice atoms is related to the energy lost by
the probe by $E_n - E_{n_0} = \frac{\hbar^2}{2m} \left( k_0^2 - k^2
\right)$. In the first Born approximation, the scattering cross
section separates into two factors. One of which is determined by the
binary interaction between the probe and each target. The other is
determined solely by the structure of the
target~\cite{vanHove1954}. This is true for general choices of the
interaction potential. In the case of the pseudopotential, the
scattered wave from an individual target is a structureless spherical
wave, and the total angular cross section is given by
\begin{equation}
\label{eq:angdiffcs}
  \frac{d\sigma}{d\Omega} = a_s^2 \sum_n \sqrt{1 -
    \frac{E_n-E_{n_0}}{\hbar^2 k_o^2/(2m)}} \ 
  \left| \bra{n} \sum_{j=1}^{N} e^{i \bm{\kappa}
      \cdot \hat{\bm{r}}_j} \ket{n_o} \right|^2.
\end{equation}
The cross section vanishes when the quantity under the square root
becomes less than zero, due to energy conservation. The momentum
transfer, $\bm{\kappa}$, depends on the index, $n$, of the final state
through the dependence on the final wavenumber, $k$, of the probe. A
natural choice of units for the cross section is the square of the
scattering length, $a_s$. The quantity under the square root
diminishes the contribution of scattering into final target states at
progressively higher energy. The matrix element of the momentum boost
gives the transition amplitude between the ground and excited states
of the target due to a transfer of momentum, $\bm{\kappa}$, from the
probe. When the interaction strength between the lattice atoms, $U$,
is very large or small compared to the tunneling matrix element, $J$,
it is possible to evaluate the cross section in~(\ref{eq:angdiffcs})
analytically. At intermediate values of $U/J$, we will numerically
diagonalize the Hamiltonian to evaluate the cross section.

For very weak interactions, we may treat the target as a condensate in
the lowest energy Bloch wave, $\psi_0(\bm{r})$, of the lowest band of
the lattice. Designating the operator that creates particles in this
state by $\hat{\psi}_0^{\dagger} = \int \! d^3r \, \psi_0(\bm{r})
\hat{\psi}^{\dagger}(\bm{r})$, the ground state of the pure superfluid
target is
\begin{equation}
\ket{n_o} = \frac{1}{\sqrt{N!}} \left(
  \frac{\hat{\psi}_0^{\dagger}}{\sqrt{N_L}} \right)^N \ket{0}. 
\end{equation}
$N$ is the number of atoms in the lattice, $N_L$ is the number of
lattice sites. In order to calculate the matrix element appearing
in~(\ref{eq:angdiffcs}) for momentum transfer, $\bm{\kappa}$, to the
target, it is useful to express the many body operator in terms of the
density, $\hat{n}(\bm{r}) = \hat{\psi}^{\dagger}(\bm{r})
\hat{\psi}(\bm{r})$, using $\sum_{j=1}^{N} e^{i \bm{\kappa} \cdot
  \hat{\bm{r}}_j} = \int \! d^3r \ e^{i \bm{\kappa} \cdot \bm{r} } \,
\hat{n}(\bm{r})$.

A convenient basis for the target in the non-interacting case is a
designation of the number of atoms in each mode of the lowest band of
the lattice. Matrix elements of the form, $\bra{n} \hat{n}(\bm{r})
\ket{n_o}$, are non-zero if at most one atom in the condensate has
been excited out of the lowest energy Bloch wave. We only need to
consider final states of the form, $\ket{n} =
\hat{\psi}_{\bm{q}}^{\dagger}\ket{\xi}$, where $\ket{\xi}$ represents
$N-1$ atoms in the ground state and $\hat{\psi}_{\bm{q}}^{\dagger}$
creates particles in the $\bm{q} \neq 0$ mode. The matrix element then
takes the form
\begin{equation}
\label{eq:density-matrix-el}
\bra{n} \hat{n}(\bm{r}) \ket{n_o} =
\frac{\sqrt{N}}{N_L} \psi^*_{\bm{q}}(\bm{r}) \psi_0(\bm{r})
\end{equation}
The energy difference of the target state, when only a single atom is
excited, reduces to the single particle energy difference between the
two Bloch waves, $E_n-E_{n_0} = \varepsilon(\bm{q}) - \varepsilon(0)$,
where $\varepsilon(\bm{q})$ is the energy of a single atom in the
specified Bloch mode of the lattice, $\bm{q}$, corresponding to the
excitation. The $\ket{n} = \ket{n_o}$ case must be handled
separately. The result we obtain for the diagonal matrix element is
\begin{equation}
\bra{n_o} \hat{n}(\bm{r}) \ket{n_o} = \frac{N}{N_L} \left|
  \psi_{0}(\bm{r}) \right|^2.
\end{equation}

We may now write down an expression for the cross section in
(\ref{eq:angdiffcs}) in the case of a superfluid ground state using
the above results for the matrix element of the density operator,
\begin{multline}
\label{eq:sfcs_Bloch_waves}
\frac{1}{a_s^2}\frac{d\sigma}{d\Omega} = N(N-1) \left| \int d^3r \
    e^{i\bm{\kappa} \cdot \bm{r}} \frac{\left| \psi_0(\bm{r})
      \right|^2}{N_L} \right|^2 \\
+ N a_s^2 \sum_{\bm{q}} \sqrt{1
    - \frac{\varepsilon(\bm{q}) - \varepsilon(0)}{\hbar^2 k_o^2/2m}}
  \left| \int d^3r \ e^{i\bm{\kappa} \cdot \bm{r}}
    \frac{\psi_{\bm{q}}^*(\bm{r}) \psi_0(\bm{r})}{N_L} \right|^2.
\end{multline}
The $\bm{q}=0$ component of the cross section gives Bragg peaks due to
the coherent overlap of elastically scattered waves from each
individual lattice site. The scale of the elastic scattering that
gives rise to the Bragg peaks can be determined by considering
scattering in the forward direction ($\bm{\kappa} = 0$). There we find
that the central Bragg peak has a height $\frac{1}{a_s^2}
\frac{d\sigma}{d\Omega} = N^2$.

The sum in the second term is over the modes of the lowest band of the
lattice. This term is exactly $N$ times the single target cross
section. This includes the elastic single target channel, $\bm{q} =
0$. We can estimate the scale of this term by considering the case in
which the probe energy significantly exceeds the bandwidth of the
lowest band, but is insufficient to excite atoms into higher
bands. The energy of the probe and depth of the lattice are
conveniently specified in units of the recoil energy, $E_r = \hbar^2
k_L^2 /(2m_T)$, for photons with wavenumber $k_L$ and lattice atoms of
mass, $m_T$. This condition is readily achieved for a typical lattice
depth of $V_o=15 E_r$, in which the width of the lowest band is $0.03
E_r$ and the band gap between the first and second bands is $6.28
E_r$. When we approximate the final wavenumber of the probe to be equal
to the initial wavenumber, the second term
in~(\ref{eq:sfcs_Bloch_waves}) becomes $N a_s^2$.

\begin{figure}[h]
\includegraphics{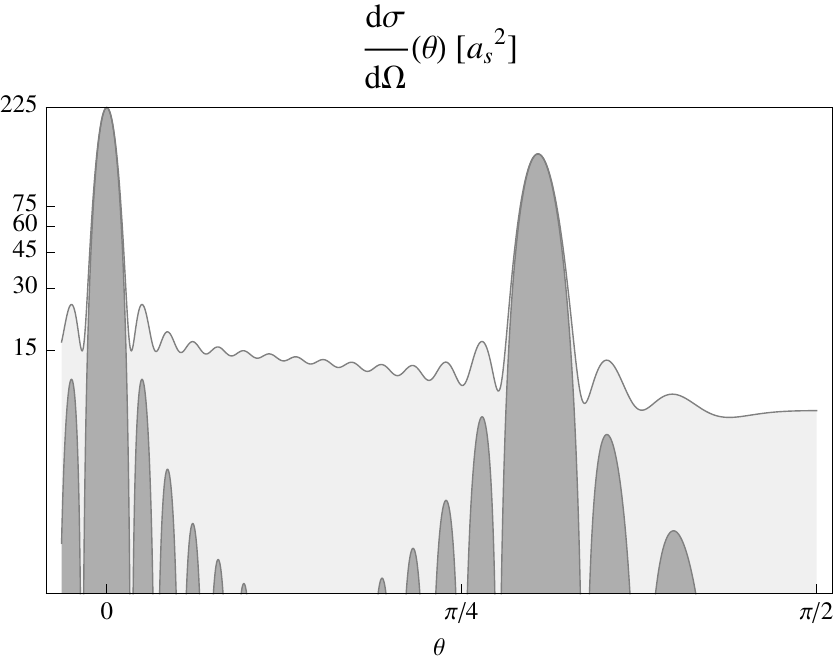}
\caption{\label{fig:anal-cs} The analytic results for the superfluid
  (light gray) and Mott insulator (dark gray) differential cross
  sections for a sample lattice in 1D with 15 atoms and 15
  sites. These are shown on a log scale in order to draw attention to
  the absence of inelastic scattering from the Mott insulator. The
  cross section is for a probe with energy $6 E_r$ and a lattice depth
  of $V_o/E_r = 15$.}
\end{figure}
There are two major features to the scattering from a superfluid:
narrow elastic Bragg peaks that scale as $N^2$ and a superimposed
inelastic background that scales as $N$. Fig.~\ref{fig:anal-cs}, which
shows the differential cross sections for the superfluid (light gray)
and the Mott insulator (dark gray), illustrates this behavior for a
one-dimensional lattice arranged perpendicular to the incident probe
wavevector. We consider the angle of deviation in the plane of the
lattice. As the number of lattice sites increases, the width of the
elastic peaks will become increasingly narrow. Away from the sharp
Bragg peaks, the inelastic background is readily identifiable. This
background emerges due to the availability of excited state modes to
the condensate. The scattering behavior is qualitatively different
when the interaction strength between atoms in the lattice becomes
very large.

As the atoms in the lattice repel each other more strongly ($U/J
\rightarrow \infty$), the superfluid fraction decreases, and the atoms
become localized within individual wells of the lattice. For a
sufficiently strong interaction, the Mott insulator state forms, and
the ground state of the target can be represented by the number of
atoms at each lattice site, $\ket{n_o} = \ket{\bar{n}_{\bm{R}_1}, ...,
  \bar{n}_{\bm{R}_{N_L}}}$. A uniform lattice will have $\bar{n} =
N/N_L$ atoms per site for integer $\bar{n}$. We must determine the
matrix element $\bra{n} \hat{n}(\bm{r}) \ket{n_o}$ for this ground
state. This is most easily done by expanding the density operator in a
Wannier basis, using the Wannier function for the lowest band of the
lattice, $w(\bm{r})$. Then $\hat{n}(\bm{r}) = \sum_{\bm{R}_1,
  \bm{R}_2} w^*(\bm{r}-\bm{R}_1) w(\bm{r}-\bm{R}_2)
\hat{a}^{\dagger}_{\bm{R}_1} \hat{a}_{\bm{R}_2}$. Each term in the sum
gives the contribution of the process in which a single target atom is
scattered from one site to another by the probe. The $\bm{R}_1 =
\bm{R}_2$ term is the elastic channel in which the state of the target
is unchanged, $\ket{n} = \ket{n_o}$. Inelastic scattering corresponds
to $\bm{R}_1 \neq \bm{R}_2$. The matrix element of the final state of
the target, in which one atom has been displaced from
$\bm{R^{\prime}}$ to $\bm{R}$, is given by \mbox{$\bra{n}
  \hat{n}(\bm{r}) \ket{n_o} = \sqrt{(\bar{n}+1) \bar{n}} \,
  w^*(\bm{r}-\bm{R}) w(\bm{r}-\bm{R}^{\prime}).$} The energy cost
associated with displacing one atom from the uniform ground state is
the interaction strength, $U$, which is also the size of the Mott
insulator gap.  These results permit us to write the explicit
expression for the scattering cross section of the Mott insulator
target,
\begin{multline}
\label{eq:MI-cs}
  \frac{1}{a_s^2} \frac{d\sigma}{d\Omega} = \bar{n}^2
  \left| \sum_{\bm{R}} e^{i \bm{\kappa} \cdot \bm{R}}
  \right|^2  \times \left| \int \! d^3r \ e^{i
      \bm{\kappa} \cdot \bm{r}}
    \left|w(\bm{r}) \right|^2 \right|^2 + \\
\bar{n}(\bar{n}+1) \sqrt{1-\frac{U}{\hbar^2 k_o^2/2m}} \\
    \times \sum_{\bm{R} \neq \bm{R}^{\prime}} \left| \int \! d^3r \
      e^{i\bm{\kappa} \cdot \bm{r}} w^*(\bm{r}-\bm{R})
    w(\bm{r}-\bm{R}^{\prime}) \right|^2.
\end{multline}
The sum over lattice sites in the first term takes a maximum value
when the momentum transferred from the probe is a reciprocal lattice
vector, with $\bm{\kappa} \cdot \bm{R}$ an integer multiple of $2
\pi$.  As the number of lattice sites increases, the elastic Bragg
peaks become increasingly sharp. In addition, there is an
approximately Gaussian envelope due to the Fourier transform of the
Wannier function.

We can examine the $\bm{\kappa}=0$ case of forward scattering, as we
did for the superfluid cross section. We see that the central peak has
a height $\frac{1}{a_s^2} \frac{d\sigma}{d\Omega} = N^2$. The elastic
scattering from the Mott insulator overlaps strongly with the elastic
scattering given by the superfluid; however, inelastic scattering from
the Mott insulator phase is strongly suppressed (see
Fig.~\ref{fig:anal-cs}). In particular, if the incident energy of the
probe is less than the Mott insulator gap, the inelastic scattering
vanishes completely. For probe energies exceeding the gap, the
inelastic scattering is also negligible under the tight binding
approximation. In that case, the integral in the inelastic part of the
cross section in~(\ref{eq:MI-cs}) is negligible when $\bm{R} \neq
\bm{R^{\prime}}$, so that we expect only elastic scattering from the
Mott insulator.

We can estimate the scale of the Mott insulator's inelastic background
more precisely by using the harmonic approximation of the Wannier
function, in which we substitute the ground state of the harmonic
approximation to the bottom of an individual well in the lattice. 
Near to the central peak, the inelastic background decays
exponentially with the lattice strength, as $\exp\left(
  -\frac{\pi^2}{2} \sqrt{\frac{V_o}{E_r}} (j-l)^2 \right)$, where $j$
and $l$ ($j\neq l$) are positions of lattice sites in units of lattice
spacings. For lattice depths in the typical range $V_o \leq 30 E_r$,
the inelastic contribution is strongly suppressed even for incident
probe energies that exceed the gap energy. This contrasts markedly
with the superfluid cross section, which carries a prominent inelastic
background that scales as the number of atoms in the lattice. In the
regions between the Bragg peaks, this background serves as an
unambiguous indicator of the many body phase of the lattice.

\begin{figure}[t]
\includegraphics[width=0.4\textwidth]{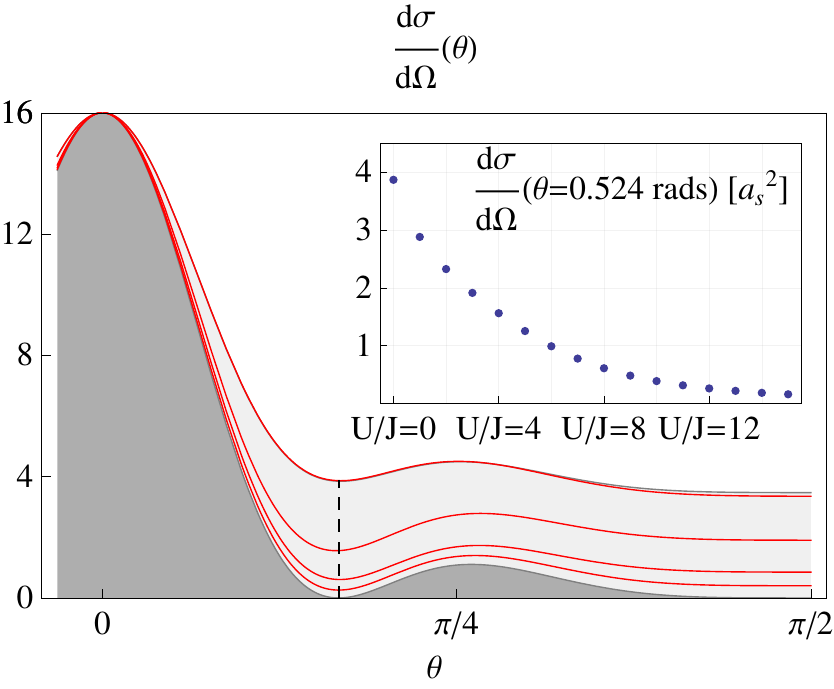}
\caption{\label{fig:num-cs} Cross sections for a probe with initial
  energy $(\hbar k_o)^2/(2m) = E_r$, a lattice of depth $V_o = 15
  E_r$, with 4 atoms and 4 sites, and $J = 0.002$. The values of $U/J$
  shown are: 0, 4, 8, and 16. The superfluid analytic result (light
  gray) for the inelastic background scales as $N$. The corresponding
  background for the Mott insulator (dark gray) is strongly
  suppressed. The inset shows the numerical cross section at the
  vertical dashed line as a function of $U/J$.}
\end{figure}

We have also examined the disappearance of the superfluid inelastic
background as the interaction strength between the lattice atoms is
increased. This required that we calculate scattering cross sections
for arbitrary values of the parameter $U/J$. At intermediate values,
this requires knowledge of the spectrum of the Bose Hubbard
Hamiltonian. We calculated the matrix elements of the density operator
by exactly diagonalizing the Bose Hubbard Hamiltonian for small
lattices. Using these results, we have shown the angular cross section
given in~(\ref{eq:angdiffcs}) for several values of $U/J$ in
Fig.~\ref{fig:num-cs}. At $U/J=0$ the numerical result coincides with
the analytic result we presented. The inelastic background that scales
as the number of atoms is present. This inelastic background decays to
zero as the interaction strength is increased, and the cross section
converges on the analytic result we gave for the Mott insulator.
We note that the amplitude of the inelastic background has decayed by
more than half at $U/J=4$. For $U/J=8$, the background is largely
gone. This coincides with the range over which the superfluid fraction
vanishes~\cite{Burnett2003}.

Our analytic results for the scattering cross section when the
interaction strength dominates the tunneling matrix element ($U \gg
J$) and vice versa show that the many body phase in the lattice
strongly affects the scattering cross section of a low-energy matter
wave probe. The periodic nature of the target gives rise in both many
body phases to coherent Bragg peaks whose height scales as the square
of the number of atoms in the lattice. In addition, an inelastic
background determined by the excitation spectrum of the target serves
as an indicator of the presence of superfluidity, and scales as the
number of atoms in the target. This provides an easily identifiable
signature of the many body phase.

Matter waves offer an effective method for non-destructively probing
the many body phase in an optical lattice. More broadly, the
scattering of matter waves depends strongly on properties such as the
distribution of atoms in a partially filled lattice, and on the long
range correlations and other manifestly quantum mechanical effects
found in this novel state of matter. This makes this technique
well-suited for examining a range of phenomena, including the effect
of inhomogeneity of bosons and fermions and disorder in the lattice,
in addition to probing the many body quantum phase transition.

\bibliography{bibliography}

\begin{thebibliography}{11}
\expandafter\ifx\csname natexlab\endcsname\relax\def\natexlab#1{#1}\fi
\expandafter\ifx\csname bibnamefont\endcsname\relax
  \def\bibnamefont#1{#1}\fi
\expandafter\ifx\csname bibfnamefont\endcsname\relax
  \def\bibfnamefont#1{#1}\fi
\expandafter\ifx\csname citenamefont\endcsname\relax
  \def\citenamefont#1{#1}\fi
\expandafter\ifx\csname url\endcsname\relax
  \def\url#1{\texttt{#1}}\fi
\expandafter\ifx\csname urlprefix\endcsname\relax\def\urlprefix{URL }\fi
\providecommand{\bibinfo}[2]{#2}
\providecommand{\eprint}[2][]{\url{#2}}

\bibitem[{\citenamefont{Jaksch et~al.}(1998)\citenamefont{Jaksch, Bruder,
  Cirac, Gardiner, and Zoller}}]{Zoller1998}
\bibinfo{author}{\bibfnamefont{D.}~\bibnamefont{Jaksch}},
  \bibinfo{author}{\bibfnamefont{C.}~\bibnamefont{Bruder}},
  \bibinfo{author}{\bibfnamefont{J.}~\bibnamefont{Cirac}},
  \bibinfo{author}{\bibfnamefont{C.}~\bibnamefont{Gardiner}}, \bibnamefont{and}
  \bibinfo{author}{\bibfnamefont{P.}~\bibnamefont{Zoller}},
  \bibinfo{journal}{Phys. Rev. Lett.} \textbf{\bibinfo{volume}{81}},
  \bibinfo{pages}{3108} (\bibinfo{year}{1998}).

\bibitem[{\citenamefont{Bloch et~al.}(2008)\citenamefont{Bloch, Dalibard, and
  Zwerger}}]{Bloch2008}
\bibinfo{author}{\bibfnamefont{I.}~\bibnamefont{Bloch}},
  \bibinfo{author}{\bibfnamefont{J.}~\bibnamefont{Dalibard}}, \bibnamefont{and}
  \bibinfo{author}{\bibfnamefont{W.}~\bibnamefont{Zwerger}},
  \bibinfo{journal}{Reviews of Modern Physics} \textbf{\bibinfo{volume}{80}},
  \bibinfo{eid}{885} (pages~\bibinfo{numpages}{80}) (\bibinfo{year}{2008}),
  \urlprefix\url{http://link.aps.org/abstract/RMP/v80/p885}.

\bibitem[{\citenamefont{Yukalov}(2009)}]{Yukalov2009}
\bibinfo{author}{\bibfnamefont{V.}~\bibnamefont{Yukalov}},
  \bibinfo{journal}{Laser Physics} \textbf{\bibinfo{volume}{19}},
  \bibinfo{pages}{1} (\bibinfo{year}{2009}).

\bibitem[{\citenamefont{Greiner et~al.}(2002)\citenamefont{Greiner, Mandel,
  Esslinger, Hansch, and Bloch}}]{Bloch2002}
\bibinfo{author}{\bibfnamefont{M.}~\bibnamefont{Greiner}},
  \bibinfo{author}{\bibfnamefont{O.}~\bibnamefont{Mandel}},
  \bibinfo{author}{\bibfnamefont{T.}~\bibnamefont{Esslinger}},
  \bibinfo{author}{\bibfnamefont{T.~W.} \bibnamefont{Hansch}},
  \bibnamefont{and} \bibinfo{author}{\bibfnamefont{I.}~\bibnamefont{Bloch}},
  \bibinfo{journal}{Nature} \textbf{\bibinfo{volume}{415}}, \bibinfo{pages}{39}
  (\bibinfo{year}{2002}), \urlprefix\url{http://dx.doi.org/10.1038/415039a}.

\bibitem[{\citenamefont{Roth and Burnett}(2003)}]{Burnett2003}
\bibinfo{author}{\bibfnamefont{R.}~\bibnamefont{Roth}} \bibnamefont{and}
  \bibinfo{author}{\bibfnamefont{K.}~\bibnamefont{Burnett}},
  \bibinfo{journal}{Phys. Rev. A} \textbf{\bibinfo{volume}{68}},
  \bibinfo{pages}{023604} (\bibinfo{year}{2003}).

\bibitem[{\citenamefont{St\"oferle et~al.}(2004)\citenamefont{St\"oferle,
  Moritz, Schori, K\"ohl, and Esslinger}}]{Esslinger2004}
\bibinfo{author}{\bibfnamefont{T.}~\bibnamefont{St\"oferle}},
  \bibinfo{author}{\bibfnamefont{H.}~\bibnamefont{Moritz}},
  \bibinfo{author}{\bibfnamefont{C.}~\bibnamefont{Schori}},
  \bibinfo{author}{\bibfnamefont{M.}~\bibnamefont{K\"ohl}}, \bibnamefont{and}
  \bibinfo{author}{\bibfnamefont{T.}~\bibnamefont{Esslinger}},
  \bibinfo{journal}{Phys. Rev. Lett.} \textbf{\bibinfo{volume}{92}},
  \bibinfo{pages}{130403} (\bibinfo{year}{2004}).

\bibitem[{\citenamefont{Mekhov et~al.}(2007)\citenamefont{Mekhov, Maschler, and
  Ritsch}}]{Ritsch2007}
\bibinfo{author}{\bibfnamefont{I.~B.} \bibnamefont{Mekhov}},
  \bibinfo{author}{\bibfnamefont{C.}~\bibnamefont{Maschler}}, \bibnamefont{and}
  \bibinfo{author}{\bibfnamefont{H.}~\bibnamefont{Ritsch}},
  \bibinfo{journal}{Physical Review Letters} \textbf{\bibinfo{volume}{98}},
  \bibinfo{eid}{100402} (pages~\bibinfo{numpages}{4}) (\bibinfo{year}{2007}),
  \urlprefix\url{http://link.aps.org/abstract/PRL/v98/e100402}.

\bibitem[{\citenamefont{Rist et~al.}(2009)\citenamefont{Rist, Menotti, and
  Morigi}}]{Morigi2009}
\bibinfo{author}{\bibfnamefont{S.}~\bibnamefont{Rist}},
  \bibinfo{author}{\bibfnamefont{C.}~\bibnamefont{Menotti}}, \bibnamefont{and}
  \bibinfo{author}{\bibfnamefont{G.}~\bibnamefont{Morigi}},
  \bibinfo{journal}{ArXiv e-prints}  (\bibinfo{year}{2009}),
  \eprint{0904.0915}.

\bibitem[{\citenamefont{Blaauboer et~al.}(2001)\citenamefont{Blaauboer,
  Kurizki, and Akulin}}]{Akulin2001}
\bibinfo{author}{\bibfnamefont{M.}~\bibnamefont{Blaauboer}},
  \bibinfo{author}{\bibfnamefont{G.}~\bibnamefont{Kurizki}}, \bibnamefont{and}
  \bibinfo{author}{\bibfnamefont{V.~M.} \bibnamefont{Akulin}},
  \bibinfo{journal}{Phys. Rev. Lett.} \textbf{\bibinfo{volume}{86}},
  \bibinfo{pages}{3518} (\bibinfo{year}{2001}).

\bibitem[{\citenamefont{W\'odkiewicz}(1991)}]{Wodkiewicz1991}
\bibinfo{author}{\bibfnamefont{K.}~\bibnamefont{W\'odkiewicz}},
  \bibinfo{journal}{Phys. Rev. A} \textbf{\bibinfo{volume}{43}},
  \bibinfo{pages}{68} (\bibinfo{year}{1991}).

\bibitem[{\citenamefont{Van~Hove}(1954)}]{vanHove1954}
\bibinfo{author}{\bibfnamefont{L.}~\bibnamefont{Van~Hove}},
  \bibinfo{journal}{Phys. Rev.} \textbf{\bibinfo{volume}{95}},
  \bibinfo{pages}{249} (\bibinfo{year}{1954}).

\end{thebibliography}

\end{document}